\documentclass{webofc}
\usepackage[varg]{txfonts}   
\usepackage{epsfig}
\usepackage{wrapfig}
\usepackage{graphicx}
\usepackage{bm}
\usepackage{amssymb}
\usepackage{amsmath}
\usepackage{amsfonts}
\usepackage[dvipsnames]{xcolor}
\usepackage{subfigure}
\usepackage{dcolumn}
\usepackage[makeroom]{cancel}
\usepackage{float}

\usepackage[utf8]{inputenc}

\usepackage{amsmath}
\usepackage{amssymb}
\usepackage{amsthm}
\usepackage{amscd, amsfonts, mathrsfs}
\usepackage{cases}
\usepackage{verbatim} 
\usepackage{epsf}
\usepackage[bookmarksnumbered,pdfpagelabels=true,plainpages=false,colorlinks=true,
            linkcolor=black,citecolor=black,urlcolor=Blue]{hyperref}

\theoremstyle{plain}

\theoremstyle{definition}

\newcommand{\Ga}{\Gamma}

\newcommand{\ep}{\varepsilon}

\newcommand{\be}{\begin{equation}}
\newcommand{\ee}{\end{equation}}
\newcommand{\bea}{\begin{aligned}}
\newcommand{\eea}{\end{aligned}}

\newcommand{\bml}{\begin{subequations}}
\newcommand{\eml}{\end{subequations}}

\newcommand{\bbm}{\begin{bmatrix}}
\newcommand{\ebm}{\end{bmatrix}}
\newcommand{\bvm}{\begin{vmatrix}}
\newcommand{\evm}{\end{vmatrix}}

\begin{document}
\title{Bulk viscosity transport coefficients
in neutron star mergers}
%
%

\author{\firstname{Yumu} \lastname{Yang}\inst{1}\fnsep\thanks{\email{yumuy2@illinois.edu}} \and
        \firstname{Mauricio} \lastname{Hippert}\inst{1}\fnsep\thanks{\email{hippert@illinois.edu}} \and
        \firstname{Enrico} \lastname{Speranza}\inst{2}\fnsep\thanks{\email{enrico.speranza@cern.ch}} \and \firstname{Jorge} \lastname{Noronha}\inst{1}\fnsep\thanks{\email{jn0508@illinois.edu}}
}

\institute{Illinois Center for Advanced Studies of the Universe \& Department of Physics, University of Illinois Urbana-Champaign, Urbana, IL 61801, USA 
\and
           Theoretical Physics Department, CERN, 1211 Geneva 23, Switzerland
          }

\abstract{%
  We compute first and second-order bulk-viscous transport properties due to weak-interaction processes in $npe$ matter in the neutrino transparent regime. The transport coefficients characterize the out-of-beta-equilibrium pressure corrections, which depend on the weak-interaction rates and the equation of state. We calculate these coefficients for realistic equations of state and show they are sensitive to changes in the nuclear symmetry energy $J$ and its slope $L$.
}
\maketitle
\section{Introduction}
\label{intro}

In binary neutron mergers, local density variations can drive the system out of beta equilibrium \cite{Hammond:2021vtv, Most:2021ktk}. Under certain conditions, weak interactions are not fast enough to restore beta equilibrium, resulting in dissipative work \cite{Alford:2018lhf, Alford:2021ogv, Alford:2023gxq}. This is particularly relevant for the neutrino transparent regime of neutron star mergers, where it was argued in Ref.\ \cite{Alford:2017rxf} that bulk-viscous dissipation
may lead to damping of density oscillations in the post-merger phase.

It was shown in Ref.\ \cite{Gavassino:2023xkt} that a two-component reactive fluid mixture can be rigorously rewritten as a resummed Israel-Stewart theory \cite{MIS-6}, with transport coefficients that explicitly depend on the bulk viscous scalar correction to the pressure. In this proceedings, we follow \cite{Gavassino:2023xkt} to determine the first and second-order bulk-viscous transport coefficients for neutrino transparent $npe$ matter using some realistic equations of state (EoS). We show that the pressure correction to beta equilibrium can be nonlinear, even near equilibrium for some EoSs, necessitating a resummed bulk-viscous description. We also investigate how the nuclear symmetry energy can affect the transport coefficients. The nuclear symmetry energy, $E_{sym}$, describes the energy difference between symmetric nuclear matter and pure neutron matter. It is usually modeled as a series expansion in baryon density $n_B$, 
\be
E_{sym}(n_B) = J + \frac{L}{3}\left( \frac{n_B}{n_{\textrm{sat}}} - 1 \right) + \mathcal{O}\left[\left(1-\frac{n_B}{n_{\textrm{sat}}}\right)^2\right],
\ee
where $J$ is the symmetry energy at the nuclear saturation density $n_{\textrm{sat}} = 0.148 \text{ fm}^{-3}$, and $L$ its slope \cite{Piekarewicz:2008nh}. To explore their impact on bulk-viscous transport coefficients, we systematically vary $J$ and $L$ (within existing constraints \cite{MUSES:2023hyz}) in this work.

\section{Resummed Israel-Stewart theory from chemical imbalance}
\label{IS derivation}

We examine neutron star mergers at low enough temperatures and densities where the neutrino mean free path exceeds the star's radius \cite{Alford:2018lhf}. The system is then neutrino transparent and only contains protons ($p$), neutrons ($n$), and electrons ($e$). When this matter gets out of beta equilibrium, flavor equilibration will then occur through direct Urca processes, $n \to p + e^- + \Bar{\nu}_e$ and $p + e^- \to n + \nu_e$, as well as modified Urca processes, $n + X \to p + e^- + \Bar{\nu}_e + X$ and $p + e^- + X \to n + \nu_e + X$,  
where $X$ is a spectator nucleon, a neutron, or a proton. 

For a neutrino-transparent system, we can write the criterion of beta equilibrium using the principle of detailed balance, 
\be
\label{eq:beta_eq}
\mu_n = \mu_p + \mu_e ,
\ee
where $\mu_n$, $\mu_p$, and $\mu_e$ are the chemical potentials for neutrons, protons, and electrons. The deviation from beta equilibrium can be measured by the difference $\delta \mu = \mu_n - \mu_p - \mu_e$, where $\delta \mu = 0$ represents beta equilibrium. In binary neutron star mergers, there are density fluctuations that occur on a millisecond timescale, which is comparable to the timescales needed by weak interactions to restore beta equilibrium \cite{Alford:2018lhf, Alford:2021ogv}. Therefore, in this case, weak interactions do not have enough time to restore beta equilibrium, resulting in dissipation \cite{Camelio:2022ljs}. 

We assume the energy-momentum tensor of the system to (formally) have the form of an ideal fluid,
\be
T^{\mu\nu} = (\ep + P)u^\mu u^\nu - P g^{\mu\nu}, 
\ee
where $\ep$ is the energy density, $P$ is the total pressure, and $g^{\mu\nu}$ is the spacetime metric. Approximating the energy loss from neutrino to be zero, the equations of motion correspond to the conservation laws 
\begin{gather}
\nabla_\mu\left[(\varepsilon+P) u^\mu u^\nu - P g^{\mu\nu}\right]= 0, \qquad \nabla_\mu (n_B u^\mu) = 0,\\
u^\mu \nabla_\mu Y_e = \frac{\Gamma_e}{n_B}, 
\end{gather}
where $n_B$ is the baryon density, $Y_e$ is the electron fraction, and $\Ga_e$ is the reaction rate. Thus, $Y_e$ is not simply transported along fluid lines due to the source term $\Ga_e$ due to the weak interactions.

Following Ref.~\cite{Gavassino:2023xkt}, one can show this system of equations is exactly equivalent to a bulk viscous description, 
\begin{gather}
\nabla_\mu\left[(\varepsilon+P_{eq}+\Pi) u^\mu u^\nu - (P_{eq}+\Pi) g^{\mu\nu}\right]= 0, \qquad \nabla_\mu (n_B u^\mu) = 0,\\
\tau_\Pi u^\mu\nabla_\mu \Pi + \Pi = - \zeta \nabla_\mu u^\mu,
\end{gather}
where the bulk scalar $\Pi = P - P_{eq}$ denotes the deviations from beta equilibrium (hence, $\Pi = \Pi(\varepsilon,n_B, \delta \mu)$). Above, $\tau_\Pi$ is the relaxation time, and $\zeta$ is the bulk viscosity. What distinguishes this theory from standard Israel-Stewart theory \cite{MIS-6} is that, here, both $\tau_\Pi$ and $\zeta$ depend explicitly on the bulk scalar, i.e., $\tau_\Pi = \tau_\Pi(\varepsilon, n_B,\Pi)$ and $\zeta = \zeta(\varepsilon, n_B,\Pi)$. Therefore, these transport coefficients $\tau_\Pi$ and $\zeta$ can depend on equilibrium and out-of-equilibrium quantities (such as $\Pi$) in the neutrino transparent regime. Alternatively, one may say that viscous corrections are \emph{resummed} into the definition of the transport coefficients of this theory.

\section{Results for the transport coefficients}

\begin{figure}[h]
\centering
\includegraphics[width=6cm,clip]{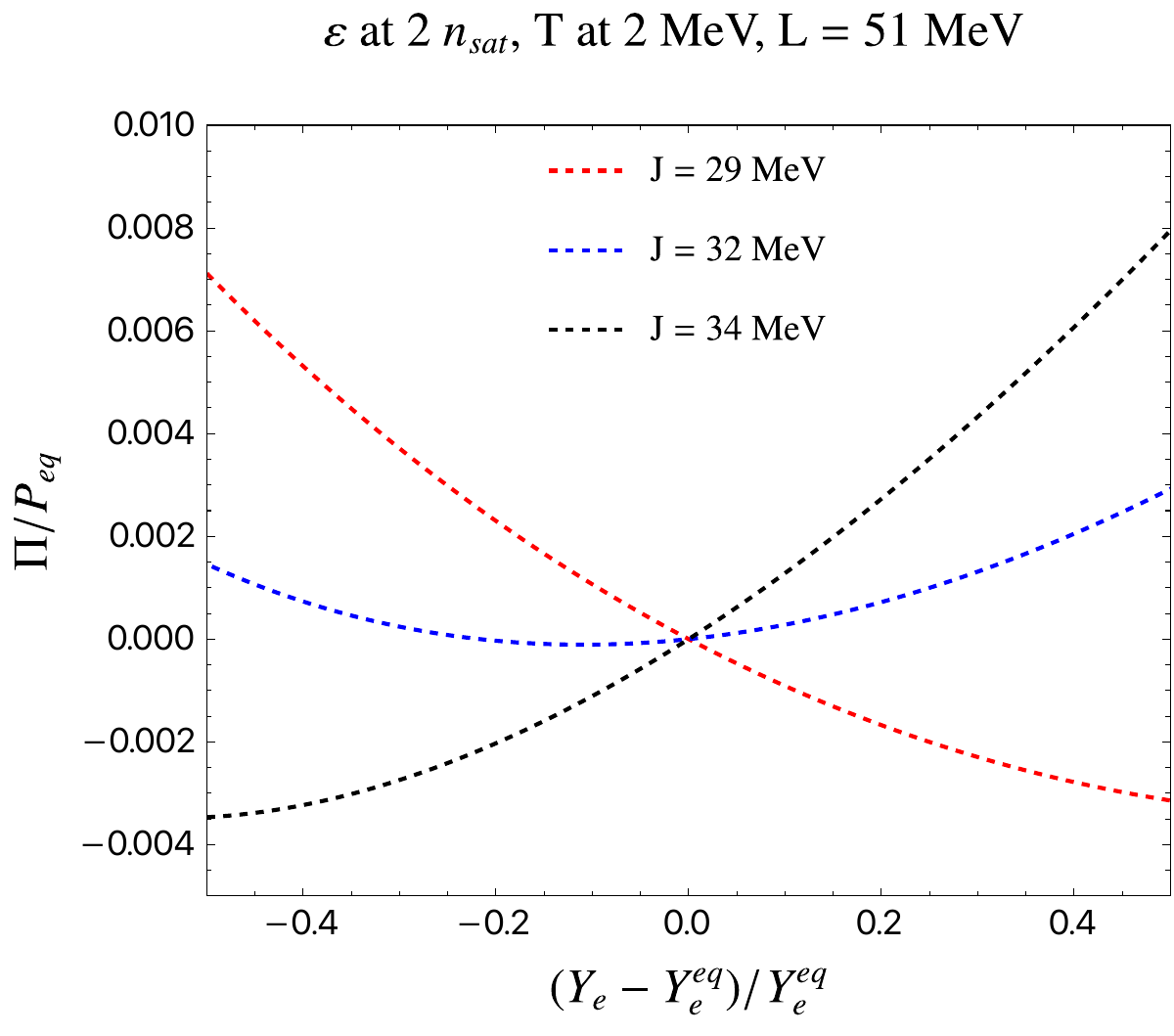}
\caption{Symmetry energy dependence of the bulk pressure as a function of the deviation from the equilibrium charge fraction, at two times saturation density, temperature $T=2$ MeV, and fixed symmetry energy slope $L$.}
\label{fig-nonlinear}       
\end{figure}

\begin{figure*}
\centering
\includegraphics[width=6cm,clip]{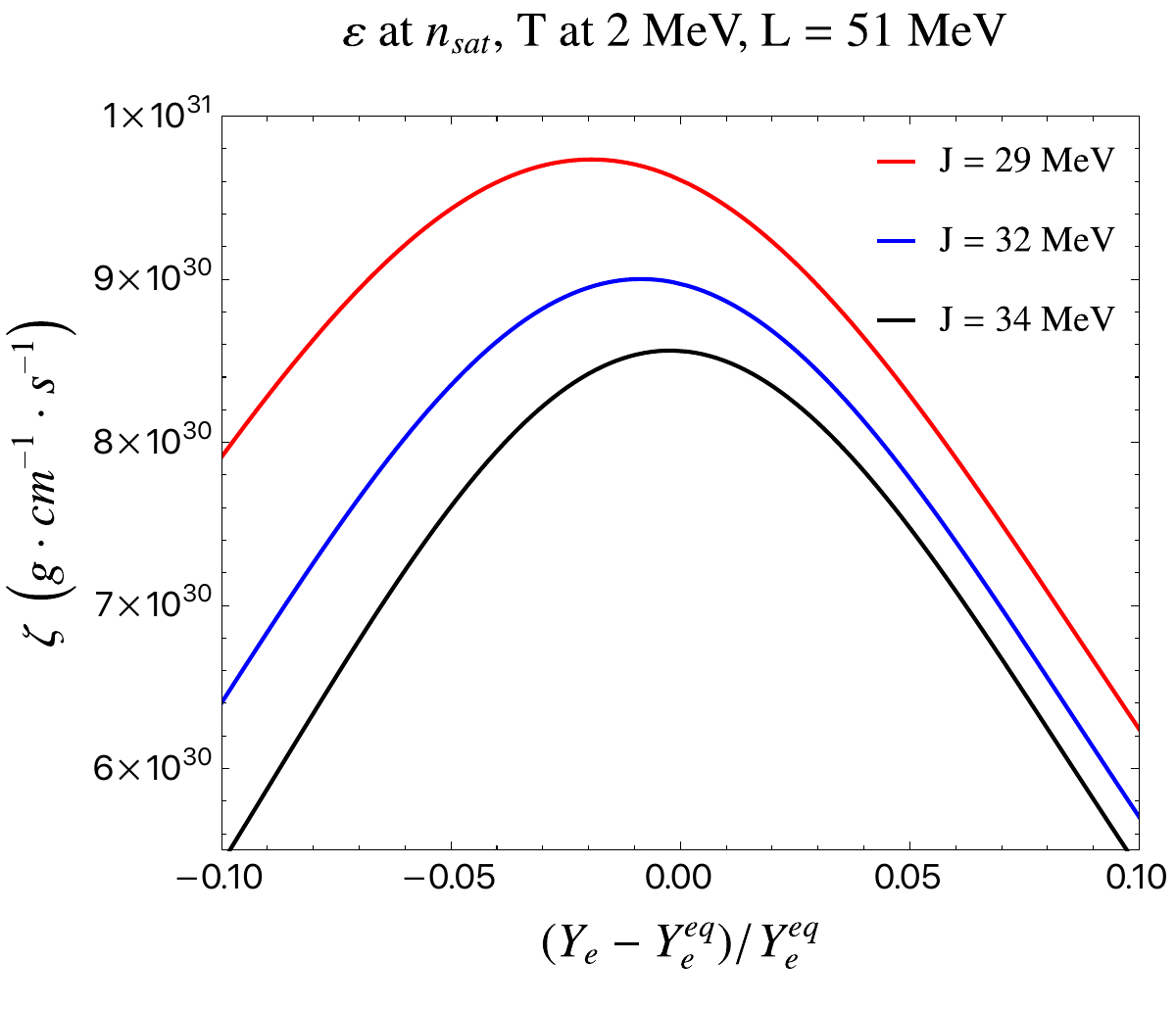} \hfill
\includegraphics[width=6cm,clip]{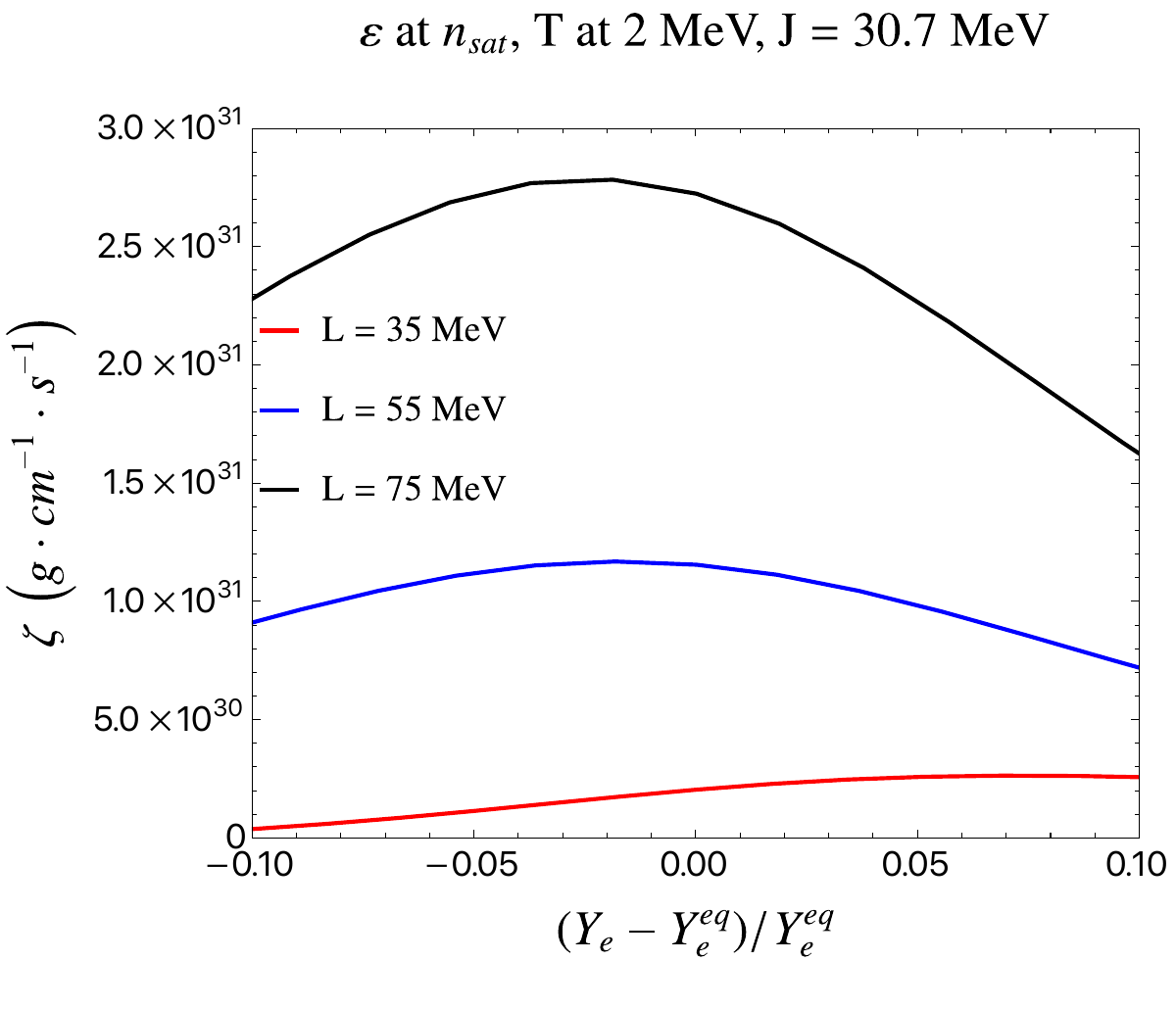} 
\caption{Resummed bulk viscosity $\zeta$ as a function of the deviation from the equilibrium charge fraction, at saturation density and temperature $T=2$ MeV. Left: Dependence on the symmetry energy at saturation, $J$. Right: Dependence on the slope of the symmetry energy, $L$. }
\label{fig-bulk}       
\end{figure*}

\begin{figure*}
\centering
\includegraphics[width=6cm,clip]{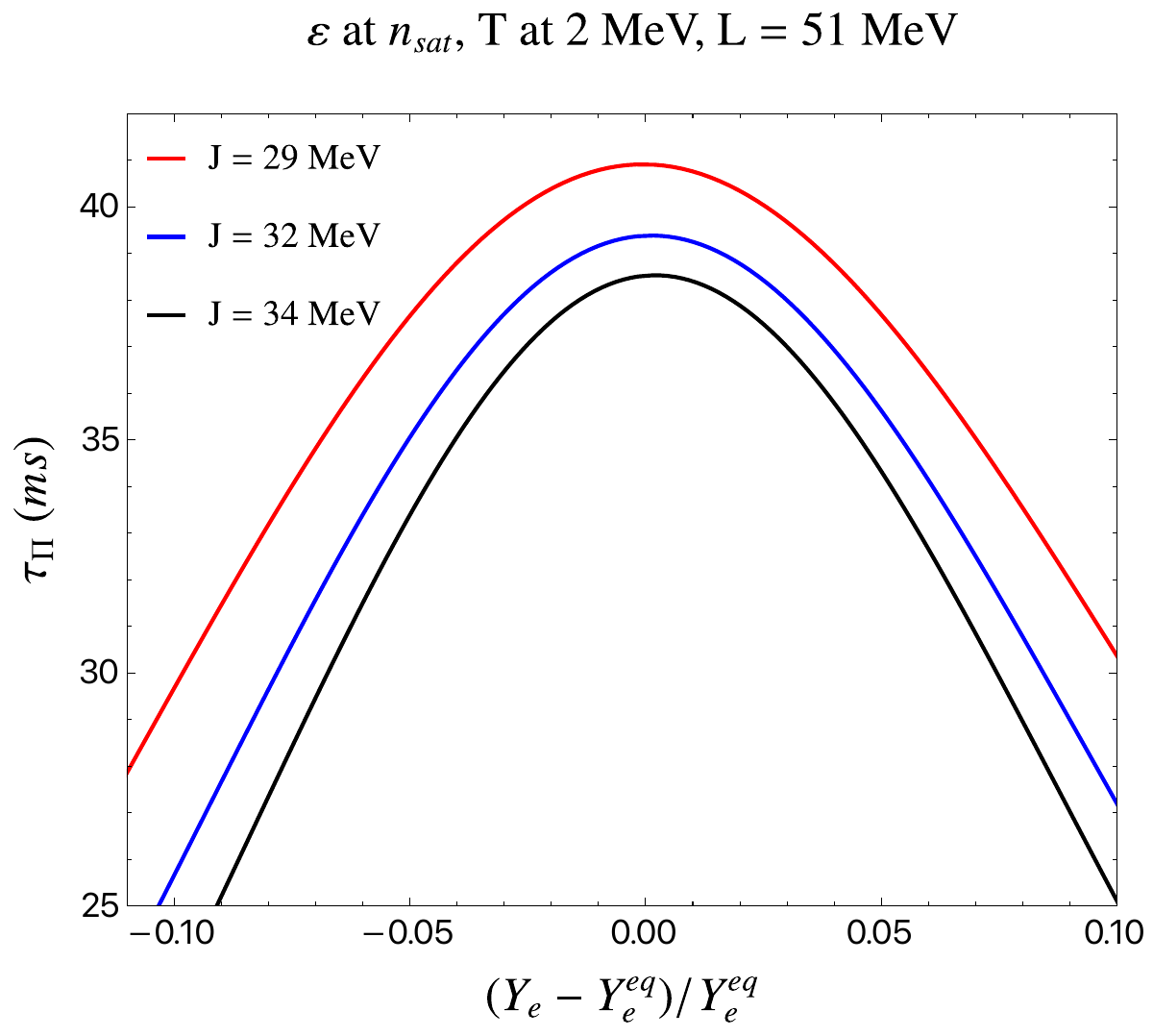} \hfill
\includegraphics[width=6cm,clip]{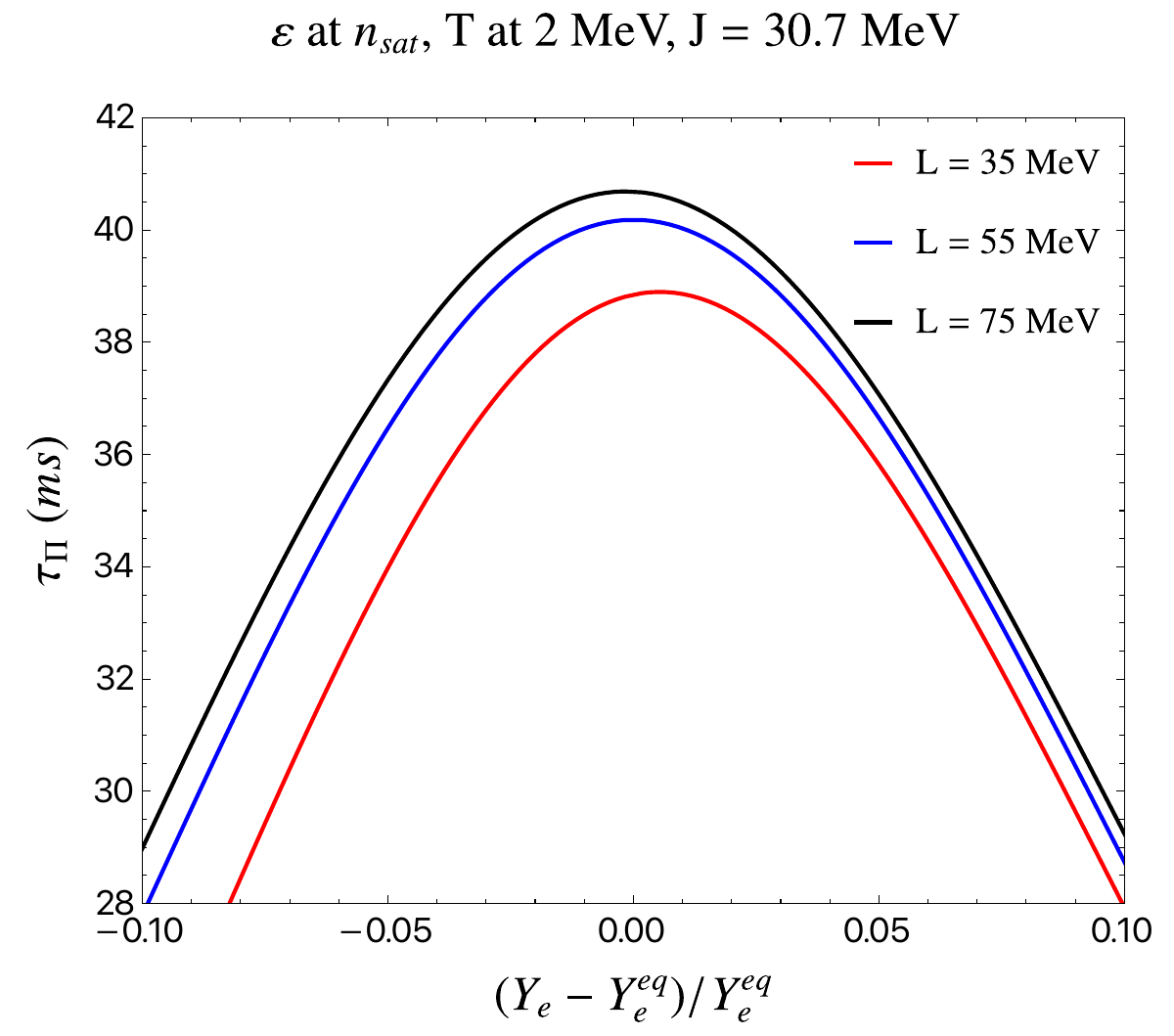} 
\caption{Resummed relaxation time $\tau_\Pi$ as a function of the deviation from the equilibrium charge fraction, at saturation density and temperature $T=2$ MeV.  Left: Dependence on the symmetry energy at saturation, $J$. Right: Dependence on the slope of the symmetry energy, $L$.}
\label{fig-time}       
\end{figure*}

We use Walecka-like relativistic mean-field models with parameters consistent with all mass-radius observations \cite{Yang:2023ogo}. As Fig.~\ref{fig-nonlinear} shows, realistic EoS can lead to nonlinear behavior of the bulk pressure near beta equilibrium. We show in Fig.~\ref{fig-bulk} and Fig.~\ref{fig-time} our results for the transport coefficients that depend on $\Pi$. One can see that the transport coefficients are quite sensitive to variations in $J$ and $L$. 

\section{Conclusion}
The neutrino transparent regime of the matter created in neutron star mergers may be described using an Israel-Stewart-like theory with transport coefficients that explicitly depend on the bulk viscous scalar. In these proceedings, we reported our results for these transport coefficients that encode far-from-equilibrium effects, showing that they are strongly affected by typical parameters of the dense matter equation of state, such as the nuclear symmetry energy and its slope. Further work is needed to determine potential gravitational-wave signatures of such bulk-viscous effects \cite{Most:2021zvc,Most:2022yhe,Chabanov:2023blf}.

\section*{Acknowledgements}
We thank J.~Noronha-Hostler, L.~Gavassino, and M.~Alford for enlightening discussions. JN is partly supported by the U.S. Department of Energy, Office of Science, Office for Nuclear Physics
under Award No. DE-SC0021301 and DE-SC002386. MH, YY, and JN were partially funded by the National Science Foundation (NSF) within the framework of the MUSES Collaboration under grant number OAC-2103680. ES has received funding from the European Union’s Horizon Europe Research and Innovation Program under grant agreement number 101109747.

\bibliography{references}

\end{document}